\def\braket#1#2#3{{\big <}#1|#2|#3{\big >}}

\def\ket#1{{\big |}#1{\big >}}
\def\smallbra#1{{<}#1{|}}
\def\smallket#1{{|}#1{>}}
\def\expect#1{{\big <}#1{\big >}}
\def\smallexpect#1{{<}#1{>}}
\def\text#1{{\mathrm{#1}}}
\def\onehalf{{\frac{1}{2}}}

\def\be{\begin{equation}}
\def\ee{\end{equation}}

\def\bea{\begin{eqnarray}}
\def\eea{\end{eqnarray}}
\documentclass[oneside,onecolumn,11pt]{article}
\usepackage{cite,epsf}

\begin{document}

\title{The AC Stark, Stern-Gerlach, \\
and Quantum Zeno Effects \\
in Interferometric Qubit Readout}


\author{J. A. Sidles
\thanks{To whom correspondence should be addressed. Email: 
\protect\texttt{sidles@u.washington.edu}}\\ 
University of Washington School of Medicine\\
Department of Orthop{\ae}dics, Box 356500\\
Seattle WA 98195}

\date{\today}

\maketitle 

\begin{abstract} 

\noindent This article describes the AC Stark, Stern-Gerlach, and
Quantum Zeno effects as they are manifested during
continuous interferometric measurement of a two-state
quantum system (qubit).  A simple yet realistic model of the
interferometric measurement process is presented, and solved
to all orders of perturbation theory in the absence of
thermal noise.  The statistical properties of the
interferometric Stern-Gerlach effect are described in 
terms of a Fokker-Plank equation, and a closed-form
expression for the Green's function of this equation is
obtained. Thermal noise is added in the
form of a externally-applied Langevin force, and the combined
effects of thermal noise and measurement are considered.
Optical Bloch equations are obtained which describe
the AC Stark and Quantum Zeno effects. Spontaneous
qubit transitions are
shown to be observationally equivalent to transitions induced
by external Langevin forces.  The effects of delayed choice
are discussed.  The results are relevant to
the design of qubit readout systems in quantum computing,
and to single-spin detection in magnetic resonance force
microscopy.

\end{abstract}
\newpage
\noindent The subject of this article is the statistical behavior of
interferometric methods for observing two-state quantum
systems (henceforth called qubits). Our analysis is
motivated by engineering and applied physics 
considerations\,---\,we wish
to determine such things as measurement times, error rates,
and whether noise-induced transitions can be suppressed by
interferometric measurement (the Quantum Zeno effect).

We shall consider interferometric measurements that are ideal
in the following respects: (1) photons are detected with
100\% efficiency, (2) the qubit-photon interaction does not
alter the helicity or wavelength of the photons, nor does it
scatter them out of the interferometer, and (3) qubit
basis states exist such that the qubit-photon interaction does
not induce transitions between the basis states.
With these restrictions, the most general scattering amplitude 
between an initial qubit state $\ket{\text{initial}}$ with an incoming photon 
state and a
final qubit state $\ket{\text{final}}$ with an outgoing photon state
is 
\be
\ket{\text{final}}\otimes 
\ket{\text{outgoing\ photon}} = 
	U \ket{\text{initial}}\otimes 
	\ket{\text{incoming\ photon}}\ ,
	\label{eq:interaction}
\ee
where $U$ is a unitary matrix acting on the two-dimensional 
qubit states.  We remark that a photon-state phase 
shift\,---\,which according to our assumptions is the only 
permissible alteration of the photon state\,---\,can always 
be absorbed into $U$, so that $U$ encompasses a completely 
general description of interferometric interactions subject 
to idealizing assumptions 1--3 above.  By an appropriate 
choice of basis, $U$ can always be written as
\be
U = \exp\left[i(\phi I + \theta \sigma_z)\right]\,,
\label{eq:phi}
\ee
where $\sigma_z$ is the usual $2\times2$ Pauli spin matrix
normalized such that $\sigma_z^2 = I$, and $I$ is the identity
matrix.  We shall see that the overall phase shift $\phi$ can 
be set to zero without loss of generality.  

It is evident that when the initial qubit state is an 
eigenstate of $\sigma_z$, the sole effect of the 
qubit-photon interaction is a state-dependent phase shift of 
the outgoing photon by $\pm\theta$, as per 
idealizing assumption~(3)~above.  We shall be particularly interested 
in measurement processes in which $\theta$ is a small 
number, $\theta \sim 10^{-6}$ or less, such that $U$ 
minimally perturbs the qubit state, as is consistent with 
real-world interferometry.

\begin{figure}[ht] 
\centerline{
\epsfxsize=5.0in
\epsfbox{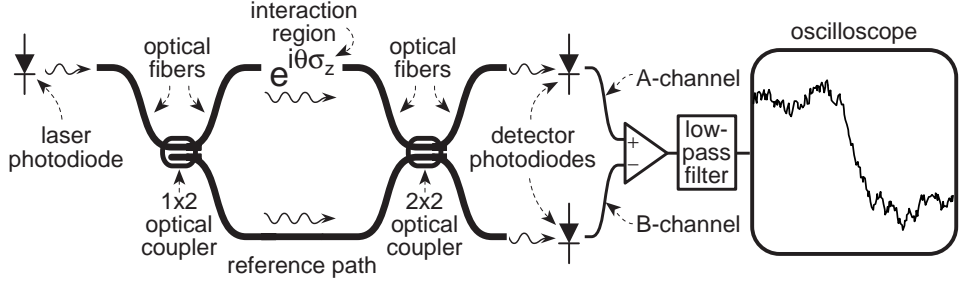}
}

\caption{Interferometer for measuring a two-state system.}
\label{fig:interferometer} 
\end{figure}

We next describe the detection of the phase shift by an 
optical interferometer of conventional design.  For 
concreteness we shall discuss a device fabricated of 
single-mode optical fiber components, as illustrated in 
Fig.~\ref{fig:interferometer}.\footnote{These optical 
components are readily 
available from commercial vendors.} Emitted photons pass 
through a beam splitter in the form of a $1\times2$ 
single-mode optical coupler.  Photons can take either of two 
paths: an upper signal path which interacts with the qubit 
system via Eq.~\ref{eq:interaction}, and a lower reference 
path which leads straight to the $2\times2$ optical coupler 
which serves as the beam combiner.  The output fibers are 
connected to conventional photodiodes, designated as the 
A-channel and B-channel in Fig.~\ref{fig:interferometer}.

As is usual in optical interferometry \cite{Rugar:89}, we stipulate that 
the reference path 
length has been adjusted such that the device is fringe-centered, 
\emph{i.e.}, for a classical phase shift $\theta$ within the
interaction region, the probability of photon detection 
on the A-channel is $(1+\sin(\theta))/2$, while on the 
B-channel it is $(1-\sin(\theta))/2$.  The effect of the 
residual phase $\phi$ in Eq.~\ref{eq:phi} is thereby
adjusted to zero.

For experiments conducted with a qubit in
the interaction region, the
data record consists of a sequence of A-channel and
B-channel detected photons: $\{a,b,a,b,\ldots ,b,a,b\}$. 
Given a starting qubit state $\ket{\text{initial}}$ and a
measured data sequence $\{a,b,a,b,\ldots ,b,a,b\}$, we 
ask: what is the qubit state $\ket{\text{final}}$ at
the end of the experiment?  For the interferometer described
above it is readily shown from Eqs.~\ref{eq:interaction}
and~\ref{eq:phi} that
the transition matrix is
\be
	\ket{\text{final}} = 
	   ABAB\cdots BAB\ \ket{\text{initial}}\,,
	   \label{eq:transition}
\ee
where the $A$ and $B$ matrices are
\bea
	A & = & \onehalf\ \left( (\cos(\theta)+i)I + 
		\sin(\theta) \sigma_z \right)\,, \\
	B & = & \onehalf\ \left( (\cos(\theta)-i)I + 
		\sin(\theta) \sigma_z \right)\,.
	\label{eq:ABdef}
\eea
The qubit state $\ket{\text{final}}$ as defined by Eq.~\ref{eq:transition}
is unnormalized; the probability of measuring a specified
data sequence $P(\{a,b,a,b,\ldots ,b,a,b\})$ is therefore
\bea
	\nonumber
	P(\{a,b,a,b,\ldots ,b,a,b\}) &=& \expect{\text{final}|\text{final}}\\
	&=& \left|\,ABAB\cdots BAB\ket{\text{initial}}\,\right|^{2}\,.
	\label{eq:prob}
\eea 
The $A$ and $B$ matrices satisfy $A^{\dagger}A + B^{\dagger}B
= I$; this ensures that the probability of each possible
data sequence, summed over all possible data sequences, is
unity.

Equations \ref{eq:interaction}--\ref{eq:prob} are the 
starting point of our analysis.  It should be noted that 
this system of equations can be regarded as a specific 
example of a well-accepted general formalism in quantum 
measurement theory\,---\,extensively discussed by, 
\emph{e.g.}, Barchielli and Belavkin 
\cite{Barchielli:91}\,---\,in which the final state 
associated with each possible data record is constructed \emph{a 
posteriori}.

Because $[A,B]=0$\,---\,later on we will discuss experiments
involving control feedback which do not satisfy this
condition\,---\,$P$ can be written in a particularly simple
form:
\bea
	P(\{a,b,a,b,\ldots ,b,a,b\}) &=&
	    \braket{\text{initial}}
	    {(A^{\dagger}A)^{n_{a}}\, (B^{\dagger}B)^{n_{b}}}
	    {\text{initial}} \label{eq:AB} \nonumber\\
	&=&  P(n_{a},n_{b})\,.\label{eq:nocollapse}
\eea
Here $n_{a}$ and $n_{b}$ are the numbers of photons measured
on the A-channel and B-channel respectively.  

It is physically significant that $P$ depends only on $n_a$ 
and $n_b$, and not on the order in which the A-channel and
B-channel photons are 
detected.  In consequence, any experimental record can be 
randomly permuted to obtain an equally probable record; thus 
the early portions of the data record are statistically 
indistinguishable from later portions, and it is impossible, 
even in principle, to identify any moment during the 
measurement process at which the qubit wave function 
collapses, even from a retrospective review of an entire 
experimental record.

It is convenient to parametrize $P(n_{a},n_{b})$ in terms of 
a single variable $q \equiv 
(n_{a}-n_{b})/((n_{a}+n_{b})\theta)$.  Physically, $q$ is 
the differential photodiode charge, with 
gain adjusted (with foresight) such that $q = \pm 
1$ is the expected value in the presence of the Stern-Gerlach effect.  
To characterize the initial qubit states, we define a qubit 
polarization variable $z_0$:
\be
z_0 \equiv 
\braket{\text{initial}}{\sigma_z}{\text{initial}}\,.
\ee
Substituting the explicit forms 
of $A$ and $B$ in Eq.~\ref{eq:AB}, and changing variables, we 
obtain the probability density $P(q |z_{0},t)$ for 
measuring $q$ after time $t$:
\be
	P(q | z_{0},t) = 
		\left[\frac{t r \theta^2}{2 \pi}\right]^{1/2} \left[\frac{(1+z_0)}{2}\,
			e^{-t r \theta^2 (q - 1)^{2}/2} + 
			\frac{(1-z_0)}{2}\, e^{-t r \theta^2 (q + 
			1)^{2}/2}\right]\,.
			\label{eq:SG}
\ee
Here $r$ is the photon flux in photons/second, and 
terms of order $\theta^3$ and higher have been dropped.  These
terms are negligible in practice, since $\theta \sim 10^{-6}$ or 
smaller is typical of optical interferometry. 

We note that $P(q | z_{0},t)$ can be written in a form in
which $z_{0}$ enters only as an overall statistical weight:
\be
	P(q | z_{0},t) = 
		\frac{(1+z_0)}{2} P(q |\,1,t) + 
		\frac{(1-z_0)}{2} P(q |-\!\!1,t)\,. \label{eq:binary}
\ee
This result implies that it is impossible, even in 
principle, to experimentally 
distinguish between (1)~a succession of experiments, each 
having identical starting polarization $z_0 = \beta$, and 
(2)~a succession of experiments, each having randomly 
assigned starting polarizations of $z_0 = \pm 1$, such that 
$z_0 = 1$ with probability $(1+\beta)/2$ and $z_0 = -1$ 
with probability $(1-\beta)/2$.

In summary, the nature of interferometric qubit measurements 
is such that, for sufficiently long measurement times, the 
measured photodiode charge~$q$ is always either $+1$ or $-1$ 
(Eq.~\ref{eq:SG})\,---\,this is the Stern-Gerlach effect.  
However, even retrospective review of the entire 
experimental record cannot identify any moment at which the 
qubit wave function collapses (Eq.~\ref{eq:nocollapse}).  
Furthermore, interferometric measurements cannot distinguish 
an ensemble of identical qubit states from a randomly mixed 
ensemble of differing states with the same average value of 
$z_0$ (Eq.~\ref{eq:binary}).  These findings collectively 
imply that interferometric measurement obeys the general 
quantum-mechanical principle that at most one bit of 
information can be obtained from measurement of a two-state 
system \cite{Wootters:82}.

Having characterized the statistical evolution of the 
\emph{macroscopic} photodiode charge~$q$ in complete detail, 
we now describe the evolution of the \emph{microscopic} 
qubit polarization variable $z$:
\be 
z \equiv \braket{\text{final}}{\sigma_z}{\text{final}}\,, 
\label{eq:zdef}
\ee 
where we specify that $\ket{\text{final}}$ (obtained from 
Eq.~\ref{eq:transition}) is to be normalized prior to 
computing $z$.  From the expressions for $A$ and $B$ 
(Eq.~\ref{eq:ABdef}) it can be shown that that $z$ evolves 
according to $dz = (1-z^{2}) \theta\, (dn_{a}-dn_{b})$.  
Then from $P(q | z_{0},t)$ (Eq.~\ref{eq:SG}) we can readily 
infer a closed-form expression for the distribution 
$P(z|z_0,t)$ which describes the statistical evolution of 
$z$:
\bea
    P(z |z_0,t)& =&  \frac{1}{1-z^{2}}\left(
	\frac{1-z_0^2}{8 \pi t r \theta^2}\right)^{1/2} 
	\left( \left(\frac{1+z}{1-z}\right)^{1/2}+
	\left( \frac{1-z}{1+z}\right)^{1/2} \right) \nonumber\\
&&\times \exp\left[\frac{-1}{8 t r \theta^2}
	\left(\ln\left(\frac{1+z}{1-z}\right)-
	\left(\frac{1+z_{0}}{1-z_{0}}\right)\right)^{2} - 
	\frac{t r \theta^2}{2}\right].
\label{eq:Greens}
\eea
By direct substitution we verify that this unwieldy
expression is the Green's function of a simple and
physically illuminating Fokker-Planck equation:
\be
	\frac{\partial P}{\partial t} = 
		r \theta^2\,\frac{\partial}{\partial z} 
		\left( D(z)\frac{\partial P}{\partial z}-V(z) P\right)\,.
		\label{eq:FokkerPlanck}
\ee
Here $D(z) \equiv (1-z^2)^2/2$ plays the role of a diffusion coefficient,
and $V(z) \equiv 2 z (1-z^2) = -dD(z)/dz$ is a drift
velocity.  

This equation has a ready physical interpretation.  From
Eqs.~\ref{eq:transition} and~\ref{eq:zdef} it can be shown that 
the microscopic qubit polarization $z(t)$ evolves as a Markovian random
process.  Qubits with $z\sim 0$ experience
maximal diffusion $D(z)$: they rapidly diffuse away from
$z\sim 0$.  Once diffusion has begun, it is hastened by the
drift velocity $V(z)$, which acts to push qubits away from
$z\sim 0$ and toward $|z|\sim 1$.  As qubit states approach
extremal values of $|z|\sim 1$, both $D(z)$ and $V(z)$ go to
zero.  In the absence of thermal noise or spontaneous
transitions\,---\,we consider these phenomena later
on\,---\,a qubit state never evolves away from the limiting
values $z = \pm1$, as shown by the asymptotic form of the
Green's function:
\be
	P(z |z_0,t) \stackrel{t\rightarrow\infty}{=}
	\frac{(1+z_0)}{2} \delta(z-1) + 
		\frac{(1-z_0)}{2} \delta(z+1)\,.
\ee
This is the Stern-Gerlach effect as manifest in the evolution of 
the quantum variable $z$; it is the microscopic analog of 
Eq.~\ref{eq:binary}.

Quantum computer designers need to take into account the
possibility that a qubit will fail to ``make up 
its mind'' during the measurement process.  At large but 
finite times we have for $|z| \not \simeq 1$
\be
	P(z |z_0,t)
	\propto \frac{1}{\sqrt{t r 
	\theta^2}} 
	\ \exp[-t r \theta^2/2]\,
\ee
Thus the error probability of interferometric qubit readout
decreases exponentially with readout time.

For the benefit of students, we remark that the Fokker-Planck 
equation (Eq.~\ref{eq:FokkerPlanck}) need not be
guessed by inspection of
its Green's function (Eq.~\ref{eq:Greens})\,---\,although this would be a legitimate
derivation strategy\,---\,but instead can be 
directly constructed by
noticing that the sequence of operators $ABAB\cdots BAB$ in
Eq.~\ref{eq:transition} defines a Markov process in $z$. 
Standard probabilistic methods (see Weissbluth
\cite{Weissbluth:89}) then suffice to construct the
Fokker-Planck equation.  With this approach, verifying that
Eq.~\ref{eq:Greens} satisfies Eq.~\ref{eq:FokkerPlanck}
becomes a check of algebraic consistency, rather than
an inspired guess.

Finally, we remark that our results up to the present
point would be unaltered if we added a Hamiltonian of
the form
\be
	H = \frac{1}{2}\hbar \omega_{0}\sigma_{z}
\ee
to the dynamical equations of the qubit state, because 
$[H,A]=[H,B]=[H,\sigma_{z}]=0$.  Here $\omega_{0}$ is the
frequency separation of the qubit eigenstates.

Next, we add thermal noise to the system.  The qubit is
now subject to two competing random
processes: the measurement process and the noise
process.  The effects of this competition 
manifest themselves as the Quantum Zeno effect.

We specify the Hamiltonian of the system as
\be
	H_{\text{int}}(t) = \frac{1}{2}\hbar \omega_{0}\sigma_{z}
		+ \hbar\,{\big [} h_{x}(t) \sigma_{x} + 
			h_{y}(t) \sigma_{y} + h_{z}(t) \sigma_{z}{\big ]}\,.
	\label{eq:Langevin}
\ee
Here $\{h_{x}(t),h_{y}(t),h_{z}(t)\}$ are Langevin fields.  
We assume that a rotating frame can be found in which their
correlation is exponential
\be
	\expect{h_{i}(t)h_{j}(t+\tau)}_{t} = 
		\delta_{ij}\, S_{i}\,e^{-\alpha \tau}\,\alpha/2\,,
		\label{eq:correlation}
\ee
where $S_{i}$ is the noise power spectral density of the 
$i$-th Langevin field $h_{i}(t)$.  We henceforth define 
$\omega_0$ in Eq.~\ref{eq:Langevin} to be the 
frequency separation of the qubit states in the preferred 
rotating frame in which Eq.~\ref{eq:correlation} applies.

We consider first the limit in which the noise decorrelation
rate $\alpha\rightarrow\infty$ (white noise), such that
$\expect{h_{i}(t)h_{j}(t+\tau)}_{t} \rightarrow
\delta_{ij}\delta(\tau) S_{i}$. The time evolution of the
qubit is still Markovian (because white noise
fields retain no memory of past values). By standard 
methods \cite{Weissbluth:89} the
Fokker-Planck equation (Eq.~\ref{eq:FokkerPlanck}) can be
generalized to include the effects of white noise:
\be
	\frac{\partial P}{\partial t} = r \theta^{2}\ 
		\frac{\partial}{\partial z} \left( 
		D(z)\frac{\partial P}{\partial z}-V(z) P \right) +
		(S_x+S_y) \frac{\partial}{\partial z} 
			\left( (1-z^2)\frac{\partial P}{\partial z}\right)\,,
			\label{eq:FokkerPlanck2}
\ee
where $\gamma \equiv 2(S_x+S_y)$. This equation implies that
the qubit polarization expectation value $\expect{z}$
defined by
\be
	\expect{z} \equiv \int_{-1}^{1} dz\,z\,P(z |z_0,t)
	\label{eq:defz}
\ee
relaxes exponentially
\be
\frac{d}{d t}\,\expect{z} = -2(S_x+S_y)\expect{z}\,,
\label{eq:Fermi}
\ee
which follows from an integration by parts. 
This we recognize as the Fermi Golden Rule, with $S_x+S_y$
the standard expression for the rate at which noise-induced
state transitions occur. The fact that $r \theta^2$ does not
appear in the transition rate is significant: it implies that
there is no Quantum Zeno effect for transitions induced by
white noise. Soon we will rederive this result in the
context of optical Bloch equations.

Despite efforts, the author has not found a closed-form 
Green's function for Eq.~\ref{eq:FokkerPlanck2}, but the 
time-independent asymptotic solution is readily obtained:
\bea
    P(z |z_0,t) &\stackrel{t\rightarrow\infty}{=}& 
    	\frac{\gamma}{(\gamma + (1-z^2))^{2}}\ 
    	\frac{(1 + \gamma)^{3/2}}{\sqrt{1 + \gamma} + 
    	\gamma\,\mathrm{arctanh}(1/\sqrt{1 + \gamma})}\nonumber\\
    	&=&\frac{\gamma}{(1-z^2)^2} + \mathcal{O}(\gamma)\,.
\eea
Here $\gamma \equiv 2(S_x+S_y)/(r \theta^2)$.  Since we
already know that $S_x+S_y$ is the rate at which transitions
occur (from Eq.~\ref{eq:Fermi}), it follows that the velocity
with which $z(t)$ transits through regions with $|z| 
\not\simeq 1$ is 
$ (1-z^2) r\theta^2/2$.

Next, we consider the effects of finite-bandwidth noise, 
\emph{i.e.}, finite values of $\alpha$ in 
Eq.~\ref{eq:correlation}.  The dynamics of the system are no 
longer Markovian\,---\,because the ``memory'' time $1/\alpha$ 
of the Langevin fields can in principle be 
long compared to the measurement time 
$1/(r\theta^{2})$\,---\,so we turn to 
time-dependent perturbation theory.  We adopt the formalism 
and notation of Weissbluth's text \cite{Weissbluth:89}.  Let 
$\rho(t)$ be the reduced density matrix of the qubit, with 
detected photons traced over.  It is readily shown that 
$\rho(t)$ obeys \be \frac{d}{dt}\,\rho = r (A\rho 
A^{\dagger} + B\rho B^{\dagger} -\rho) - 
\frac{i}{\hbar}[H_{\text{int}}(t),\rho]\,.
		\label{eq:rhodot}
\ee
In the absence of $H_{\text{int}}$ this equation is linear
with constant coefficients, and is therefore exactly
solvable in terms of exponentials.  Perturbing about the
no-noise exact solution, and applying standard methods of
second-order time-dependent perturbation theory
\cite{Weissbluth:89}, we obtain the Green's function of
$\rho(t)$ correct to all orders in photon flux $r$ and second order in
Langevin fields $\{h_i\}$.  The resulting expression for $\rho(t)$ becomes
simple in the physically interesting limit in which
measurement occurs on time scales short compared to
noise-induced relaxation, \emph{i.e.}, $S_i \ll
r\theta^{2}$. In such experiments the evolution of $\rho(t)$
can be described compactly in terms of optical Bloch
equations. We define optical variables in the usual way
\cite{Weissbluth:89}:
\be 
	u_{x}(t)  \equiv  \text{Tr}(\sigma_{x}\rho(t))\,,\ 
	u_{y}(t)  \equiv  \text{Tr}(\sigma_{y}\rho(t))\,,\ 
	u_{z}(t)  \equiv  \text{Tr}(\sigma_{z}\rho(t))\,.
\ee
It is readily shown that $u_{z} = \expect{z}$, where \expect{z} is the 
previously-defined qubit polarization expectation value 
(Eq.~\ref{eq:defz}).\footnote{The relation is proved as follows.  We consider an 
ensemble of $n$ normalized qubit states $\{\smallket{q_{i}};i=1,n\}$, each 
with associated probability $p_{i}$ and polarization $z_{i}$.  In the limit of large $n$ we 
define a probability density $P(z)$ such that
\bea
\smallexpect{z}\hspace{-0.08in} & =  &\hspace{-0.08in} \int_{-1}^{1} dz\,z\,P(z)
= \sum_{i=1}^{n} p_{i} z_{i} = \sum_{i=1}^{n} p_{i} 
\smallexpect{q_{i}|\sigma_{z}|q_{i}}\nonumber\\
&=& \hspace{-0.08in} \text{Tr}\left(\sigma_{z}\left( \sum_{i=1}^{n}  p_{i} 
\smallket{q_{i}}\smallbra{q_{i}}\right)\right)
= \text{Tr}(\sigma_{z}\rho)\,,\nonumber
\eea
as was to be shown.} This relation will prove useful in checking the 
consistency of the optical Bloch equations with the Fokker-Planck 
equation.

It is readily demonstrated that the leading asymptotic terms of
the Green's function of Eq.~\ref{eq:rhodot}
are generated by the following optical Bloch equations:
\bea
	\label{eq:opticalBloch1}
	\frac{du_{x}}{dt}  & = & -\,\frac{u_{x}}{T_{2}} - 
		(\omega_0 - r \theta)\,u_{y}\,,\\
		\label{eq:opticalBloch2}
	\frac{du_{y}}{dt}  & = & -\,\frac{u_{y}}{T_{2}} + 
		(\omega_0 - r \theta)\,u_{x}\,,\\
		\label{eq:opticalBloch3}
	\frac{du_{z}}{dt}  & = & -\,\frac{u_{z}}{T_{1}}\,.\label{eq:uz}
\eea
The relaxation times $T_{1}$ and $T_{2}$ are found to be
\bea
\label{eq:QuantumZeno}
	\frac{1}{T_{1}} & = & 2 (S_{x}+S_{y})\ 
		\frac{\alpha (\alpha+ r\theta^{2}/2)}
		{(\alpha+ r\theta^{2}/2)^{2}+(\omega_0 - r \theta)^{2}}\,,\\
	\frac{1}{T_{2}} & = & \frac{r \theta^{2}}{2} + {\mathcal O}(S_i)\,.
	\label{eq:T1T2}
\eea
Essentially identical optical Bloch equations are obtained 
for the case of spontaneous emission into a continuum of 
vacuum states, provided we stipulate that the density of 
vacuum states $n(\omega)$ is Lorentzian, \emph{i.e.}, 
$n(\omega) \propto 1/(\omega^2+\alpha^2)$, and we substitute 
$S_{x}+S_{y}\rightarrow \Gamma$, with $\Gamma$ the 
spontaneous transition rate in the absence of measurement.  
The sole adjustment to the form of the optical Bloch 
equations is that Eq.~\ref{eq:uz} is modified to read
\be
	\frac{du_{z}}{dt} =  -\,\frac{1+u_{z}}{T_{1}}\,,
\ee
reflecting the fact that spontaneous transitions drive the 
system irreversibly toward the low-energy qubit state $u_z = 
-1$ rather than $u_z = 0$.  With these substitutions, 
Eqs.~\ref{eq:opticalBloch1}--\ref{eq:T1T2} describe both 
spontaneous and stimulated transitions.

It is apparent that in spontaneous
transitions, the frequency bandwidth of the vacuum states
plays the same role as the frequency bandwidth of the
Langevin noise in stimulated transitions.

Equations \ref{eq:opticalBloch1}--\ref{eq:T1T2} embody the 
main subjects of this article: the AC Stark, Stern-Gerlach, 
and Quantum Zeno effects.  The AC Stark effect is readily 
apparent in the optical Bloch equations for the transverse 
qubit polarization $u_x$ and $u_y$ 
(Eqs.~\ref{eq:opticalBloch1} and~\ref{eq:opticalBloch2}): 
the photon-qubit interaction shifts the optical Bloch precession 
frequency by an amount $r \theta$.  The AC Stark effect also 
appears in the relaxation rate of $u_{z}$ 
(Eq.~\ref{eq:opticalBloch3}): stimulated qubit transitions 
are on-resonance only if the Langevin carrier frequency is 
shifted by $\omega_{0} = r \theta$.  The sign of the AC 
Stark effect is such that a positive phase shift $\theta$ 
generated by the higher energy state induces reduced energy 
separation between states during measurement.  Conversely, 
negative phase shifts will induce increased energy 
separation.  The AC Stark shift must be taken into account 
in the design of experiments in which the Langevin fields 
$\{h_i(t)\}$ are externally supplied, are narrow-band, and 
are intended to be on-resonance during the measurement 
process.

The most striking aspect of the Stern-Gerlach 
effect\,---\,the measurement of $z= \pm 1$ in individual 
experiments\,---\,is not immediately apparent in the optical Bloch 
equations, which describe only the mean qubit 
polarization.  
However, there is a rigorous sense in which the optical Bloch equations still 
embody\,---\,indirectly\,---\,the Stern-Gerlach effect.
We note that the Stern-Gerlach effect implies
that, at sufficiently late times, the transverse  
polarizations $\{u_{x},u_{y}\}$ must vanish.  Building on this idea,
we shall prove that the transverse polarization
relaxes at the slowest possible rate consistent with the 
Stern-Gerlach effect.  We 
begin by noting that the optical Bloch equations 
(Eqs.~\ref{eq:opticalBloch1} and~\ref{eq:opticalBloch2})
imply that the squared transverse qubit 
polarization relaxes with relaxation rate $r \theta^2$:
\be
	\frac{d}{dt} (u_x^2 + u_y^2) = 
		-r \theta^2\,(u_x^2 + u_y^2)\,,
		\label{eq:transverse}
\ee
Similarly, we note
that the Fokker-Planck equation (Eq.~\ref{eq:FokkerPlanck}) implies that
ensemble-averaged expectation value $\expect{1-z^2}$ relaxes according to
\bea
\nonumber
	\frac{d}{dt} \expect{1-z^2} &\equiv& \frac{d}{dt}\ \int_{-1}^{1} 
	dz\,(1-z^2)\,P(z |z_0,t)\\
	&=&  -r \theta^2\,\expect{(1-z^2)^2}\,,
	\label{eq:bound}
\eea
as follows from an integration by parts.  Unfortunately, the 
squaring of the right-hand side ensures that the time 
dependance of $\expect{1-z^2}$ is more complex than a simple 
exponential; no closed-form solution of Eq.~\ref{eq:bound} 
is available.\footnote{The closely-related
expectation value $\smallexpect{\sqrt{1-z^{2}}}$ does relax 
exponentially,
with rate constant $r\theta^{2}/2$, but this does not help 
prove the desired theorem.  We note as a point of interest 
that the expectation value $\smallexpect{z/\sqrt{1-z^{2}}}$
grows exponentially, with time constant $3 r \theta^{2}/2$,
consistent with the Fokker-Planck equation driving values of
$z(t)$ towards $|z| \sim 1$.} To make progress, we note a readily-derived 
general inequality relating the optical Bloch variables 
$u_x$ and $u_y$ to the Fokker-Planck variable 
$z$:
\be 
	u_x^2 + u_y^2 \le \expect{1-z^2}\,.
	\label{eq:inequality} 
\ee 
The equality holds for pure (unmixed) qubit states.  It 
follows that the relaxation of an initially pure qubit state
must satisfy
\be
\left.\left(\frac{d}{dt} (u_x^2 + u_y^2)\right)\right|_{t=0} \le 
\left.\left(\frac{d}{dt} \expect{1-z^2}\right)\right|_{t=0}
\label{eq:constraint}
\ee  
in order that Eq.~\ref{eq:inequality} hold to first order in 
$t$ for times $t>0$.  Evaluating the left-hand side of 
Eq.~\ref{eq:constraint} via the optical Bloch equations 
(Eq.~\ref{eq:transverse}), and the right-hand side via the 
Fokker-Planck equation (Eq.~\ref{eq:bound}), and setting 
$u_x^2 + u_y^2 = \expect{1-z^2} = 1-z_0^2$ for a pure state 
with $z=z_0$ at time $t=0$, we obtain the sought-after 
inequality constraining the optical Bloch transverse 
relaxation rate relative to the Fokker-Planck relaxation 
rate:
\bea 
\nonumber
\mbox{(optical\ Bloch\ time\ derivative)} & \le & 
\mbox{(Fokker-Planck\ time\ derivative)}\,,\\
-r \theta^2 (1-z_0^2) &\le& -r \theta^2 (1-z_0^2)^2\,.
\eea
We observe that this inequality is satisfied for all values 
of initial qubit polarization $z_0$, which confirms the 
consistency of the Fokker-Planck and optical Bloch 
formalisms.  The inequality is saturated for the particular 
case $z_0 = 0$, \emph{i.e.}, purely transverse initial qubit 
polarization.  We have therefore shown that the relaxation 
of transverse polarization described by the optical Bloch 
equations takes place at the slowest possible rate 
consistent with the Stern-Gerlach effect as described by the 
Fokker-Planck equation.

Similarly, we can check the $z$-component optical Bloch
equation (Eq.~\ref{eq:opticalBloch2}) for consistency with
the Fokker-Planck equation by noting that in the white-noise
limit $\alpha\rightarrow\infty$, the optical Bloch
relaxation rate for $u_z$ is $1/T_1 \rightarrow 2
(S_{x}+S_{y})$, which is identical to the relaxation rate
for $u_z = \expect{z}$ predicted by the white-noise Fokker-Planck
equation (Eq.~\ref{eq:Fermi}).

The optical Bloch expression for $1/T_1$ (Eq.~\ref{eq:QuantumZeno})
embodies the Quantum Zeno effect. Assuming the AC Stark
effect has been tuned to zero (\emph{i.e.}, $\omega_0 = r
\theta$), continuous interferometric measurement suppresses
noise-induced relaxation by a factor
\be
\frac{1/T_1\,(\text{with\ 
measurement})}{1/T_1\,(\text{without\ 
measurement})} = \frac{\alpha}{(\alpha+ r\theta^{2}/2)}\,.
\ee
Physically speaking, the Quantum Zeno effect is observed
whenever the noise bandwidth $\alpha$ (or for the case of
spontaneous transitions, the bandwidth $\alpha$ of the
available vacuum states) is of the same order or smaller
than the measurement bandwidth $r\theta^{2}/2$. 

\begin{figure}[t] 
\centerline{
\epsfxsize=3.75in
\epsfbox{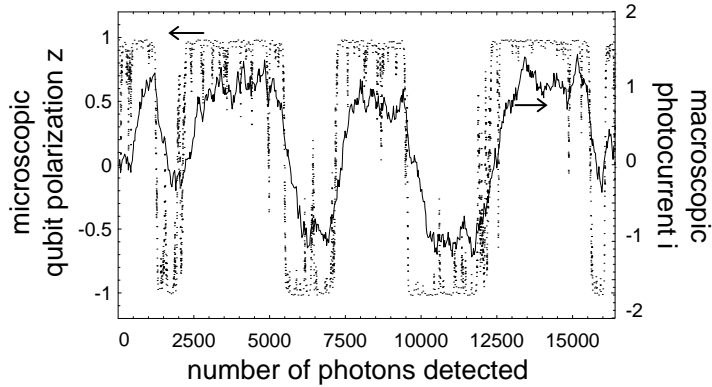}
}
\caption{Microscopic and macroscopic variables.}
\label{fig:variables} 
\end{figure}

\begin{figure}[t] 
\centerline{
\epsfxsize=3.75in
\epsfbox{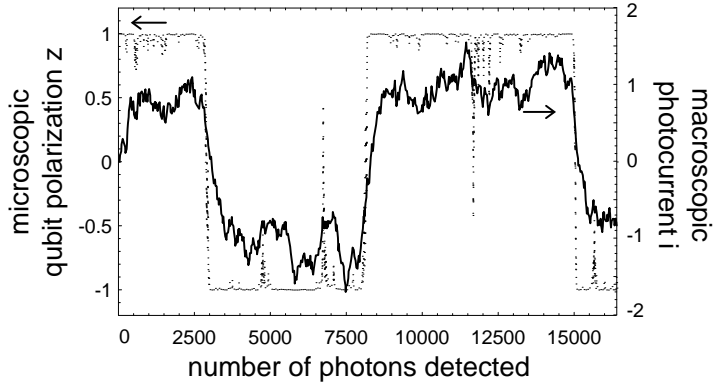}}
\caption{The Quantum Zeno effect.} 
\label{fig:Zeno} 
\end{figure}

We conclude our discussion by presenting numerical
simulations of the combined effects of measurement and noise
(Figs.~\ref{fig:variables} and~\ref{fig:Zeno}).  In both
experiments $2^{14} = 16,384$ photons are sent through the
system, with phase shift $\theta = 1/\sqrt{64}$. We adopt
units of time such that $r=1$, hence the measurement
bandwidth $r \theta^{2}/2 = 1/128$, so typically 128 photons
suffice to measure the qubit state.  The macroscopic
photodiode current $i$ shown in the figures represents the
photodiode current after low-pass
filtering: the single-pole $RC$ filter time constant has been set
to $1/(RC) = 1/512$ and the zero-frequency filter gain adjusted 
to $1/(e r \theta)$, where $e$ is the photodiode charge 
collected for each detected photon, such
that currents of $i = \pm 1$ are expected for qubit
polarization $z = \pm 1$.  The noise
fluctuations evident in the measured current are wholly due
to photodiode shot noise.  The spectral density of the
Langevin noise is adjusted such that $S_x + S_y = 1/2048$;
thus a mean of eight transitions are expected in the course
of each experiment, prior to taking the Quantum Zeno effect
into account.  

Fig.~\ref{fig:variables} illustrates a typical simulation.  
The effects of diffusion and drift velocity in the 
Fokker-Planck equation (Eq.~\ref{eq:FokkerPlanck2}) are 
readily evident in the microscopic qubit 
polarization\,---\,$z(t)$ does not linger near $z\sim 0$, 
but rather is swiftly drawn to values of $z\sim \pm 1$.  
Conversely, for values of $z(t)$ near $\pm 1$, Langevin 
noise generates numerous tunneling attempts of which only a 
small fraction succeed, in accord with both the 
Fokker-Planck and optical Bloch equations.  Transitions in 
the macroscopic photocurrent lag behind those of the 
microscopic variable $z$ due to the causal nature of the 
low-pass filter.  The transitions in 
Fig.~\ref{fig:variables} are induced by broad-band Langevin 
noise, with $\alpha = 1/16$, such that the predicted Quantum 
Zeno suppression factor is $\alpha/(\alpha+ r\theta^{2}/2) = 
8/9$\,---\,essentially negligible.

Fig.~\ref{fig:Zeno} shows a simulation in which the Langevin 
noise has the same spectral density as 
Fig.~\ref{fig:variables}, but is narrow-band, $\alpha = 
1/256$, such that the Quantum Zeno suppression factor is 
$\alpha/(\alpha+ r\theta^{2}/2) = 1/3$.  The predicted 
three-fold suppression of noise-induced transitions is 
readily evident in $z(t)$, not only in the reduced number of 
outright transitions (from eight to three), but also in the 
reduced number and intensity of tunneling attempts.

We note that the interferometric Quantum Zeno effect 
discussed in the present article differs substantially from 
the Quantum Zeno effect experimentally observed by Itano 
\emph{et.\ al.}\ \cite{Itano:90}, as reviewed by Wineland 
\emph{et.\ al.}\ \cite{Wineland:95}.  One major difference 
is that interferometric techniques probe two-state qubit 
systems, in contrast to the three-state ion systems of Itano 
\emph{et.\ al.}.  Furthermore, interferometric qubit-photon 
interactions are off-resonance and weak, and do not induce 
state transitions, while in the experiments of Itano 
\emph{et.\,al.}\ the ion-photon interactions were 
on-resonance and strong, and did induce state transitions.  
Similarly, the quantum jumps visible in 
Figs.~\ref{fig:variables} and~\ref{fig:Zeno} are visually 
similar to the three-state quantum jumps experimentally demonstrated by 
Dehmelt \cite{Dehmelt:90}, but differ substantially in their 
underlying mechanism.

The intimate relation between the AC Stark, Stern-Gerlach, 
and Quantum Zeno effects has not previously 
been described in the literature\,---\,all three effects can 
be predicted from knowledge of the photon flux $r$, the 
optical scattering phase $\theta$, and the fluctuation 
bandwidth $\alpha$, and the observation of any one effect 
implies the presence of the other two.

The results presented in this article are of substantial 
utility in the design of interferometric experiments.  As an 
example, let us design a trapped-ion experiment that would 
generate oscilloscope traces statistically identical to 
those of Figs.~\ref{fig:variables} and 
\ref{fig:Zeno}\,---\,traces in which Stern-Gerlach and 
Quantum Zeno phenomena are readily apparent to the naked 
eye.  We stipulate that the duration of the oscilloscope 
trace is to be 10 seconds, and the detected photon flux is 
$1.0\ \mu \text{W}$ at a wavelength of 6800 \AA, for a total 
of $n_{\text{exp}} = r t = 3.4 \times 10^{13}$ detected 
photons.  Only half of this optical power passes through the 
interaction region; the rest is carried by the reference arm 
of the interferometer.  The low-pass filter time constant 
should be set to 1/32 of the trace time, \emph{i.e.}, $RC = 
0.3\ \text{sec}$, in order to duplicate the shot-noise 
fluctuations evident in the oscilloscope traces of 
Figs.~\ref{fig:variables} and~\ref{fig:Zeno}.  The traces 
have a resolving power $r t \theta^{2} = 256$: the designed 
experiment is intended to achieve the same resolving power.  
This implies that the target two-state ion must 
generate an optical phase shift of at least 
$\theta_{\text{exp}} = (256/n_{\text{exp}})^{1/2} = 2.7 
\times 10^{-6}$ radians.  The required AC Stark frequency 
shift is $\delta \omega = r \theta = 2.8\ \text{MHz} = 5.8 
\times 10^{-4}\ \text{cm}^{-1}$.  If the photon flux were boosted 
by a factor of $10^{6}$ (to one Watt), the detectable phase 
shift would be reduced by a factor of $10^{-3}$ to $2.7 \times 
10^{-9}$ radians; this square-root scaling is characteristic 
of interferometric detection.  The required AC Stark shift 
would be $r \theta = 2.8\ \text{GHz} = 0.58\ \text{cm}^{-1}$.  

In practical design work, it will often be convenient to 
to regard the photon flux $r$ as the primary design 
variable, then estimate the Stark shift $\delta \omega$ from 
the optical field intensity in the interaction region and
the optical properties of the ion under investigation.  The 
scattering phase $\theta$ can then be determined simply from 
$\theta = \delta \omega / r$.

It is beyond the scope of this article to evaluate the 
photon fluxes and Stark shifts that are practically 
achievable in trapped ion experiments; our concern is to 
provide specific design targets which contemplated 
experiments must meet.

For the benefit of students new to quantum mechanics, we 
remark that introductory textbooks often contain simplified 
or axiomatic descriptions of measurement processes which 
sometimes lend an unnecessarily paradoxical aspect to 
well-understood phenomena like the Stern-Gerlach effect.  The results 
presented in this article are in accord with an increasingly 
dominant modern view\,---\,but a view requiring 
substantially more complicated calculations than are 
typically included in introductory texts\,---\,in which 
measurement processes work gently and incrementally to 
create correlations between macroscopic variables 
(like photodiode charge~$q$) and microscopic 
variables (like qubit polarization~$z$).  At the end 
of an interferometric qubit measurement, all but an 
exponentially small fraction of data records agree that the 
Stern-Gerlach effect is present, but it is both unnecessary 
and impossible, even in principle, to identify a specific 
moment at which the qubit wave function collapsed.  Students 
should also be aware that the extra work entailed in a 
detailed quantum measurement analysis can yield worthwhile 
physical insights.  For example, students may reflect on our 
finding that spontaneous emission into a broad continuum of 
states is not subject to the Quantum Zeno effect\,---\,this 
greatly simplifies the calculation of atomic transition 
rates in photon-rich environments like stellar interiors, 
where Quantum Zeno suppression would otherwise be a large 
effect.

Students should also be aware that the level of detail 
in the analysis of quantum measurement processes is limited 
mainly by space considerations and by the energies of the 
investigator and the reader.  For example, we have not 
specified what happens inside the photodiodes of 
Fig.~\ref{fig:interferometer}, but there is no reason to 
think that a more detailed description of these photodiodes 
would change our conclusions regarding the AC Stark, 
Stern-Gerlach, and Quantum Zeno effects. 

The results of this article can be extended in several 
directions.  We have considered the measurement of a single 
qubit, yet the design of practical quantum computers will 
require simultaneous interferometric measurements on 
multiple correlated qubits \cite{DiVincenzo:95}: the 
detailed theoretical analysis of this measurement process 
remains to be done.  Single-spin magnetic resonance force 
microscopy ({\small MRFM}) \cite{Sidles:91,Sidles:95} is 
based on the interferometric measurement of a harmonic 
oscillator that is coupled to a two-state spin 
system.  Stern-Gerlach phenomena are predicted to occur 
\cite{Sidles:92}, but as yet there has been no 
quantum measurement analysis of this more complex 
interferometer-oscillator-spin system.  The results 
of the present article were derived as a preliminary step 
toward a quantum measurement analysis of 
single-spin {\small MRFM}.

Qubit interferometry readily lends itself to delayed-choice 
experiments, which have yet to be analyzed in detail.  We 
imagine, for example, that the $2 \times 2$ optical coupler 
of Fig.~\ref{fig:interferometer} is separated from the 
interaction region by 150 kilometers of optical fiber, 
yielding a full millisecond of time delay in the measurement 
process\,---\,such delays are perfectly feasible from a 
technical point of view.  We launch a microsecond-long pulse 
of $n_{\text{exp}} \sim 10^{13}$ photons into the apparatus.  
After the photons have interacted with the qubit, but before 
they have entered the $2 \times 2$ coupler, we choose 
whether or not to insert the coupler in the system.  This 
delayed choice retrospectively alters the quantum state of 
the qubit, as per Eq.~\ref{eq:Greens} with 
$rt=n_{\text{exp}}$.  To avoid the possibility of a 
backwards-in-time causality violation, the alteraction of 
the qubit state consequent to the delayed choice must be 
undetectable.  In the special case that Langevin noise is 
absent, it can be shown from Eq.~\ref{eq:binary} that the 
retrospective qubit state alteraction has no detectable 
consequences, but a general proof that all delayed-choice 
qubit interferometry experiments respect causality has not 
yet been given.

Interesting phenomena\,---\,as yet poorly 
understood\,---\,occur when the condition $[A,B]=0$ is 
relaxed.  Physically speaking, this occurs when a 
feedback loop is installed, such that a unitary transform 
$U_a$ is applied to the qubit whenever an A-channel photon is 
detected, while a (possibly different) unitary transform 
$U_b$ is applied whenever a B-channel photon is detected.  Then 
Eqs.~\ref{eq:interaction}--\ref{eq:prob} still apply, with 
the substitutions \bea
\nonumber A \rightarrow  A'&\!\!\equiv\!\!&U_a A\,, \\
B \rightarrow  B'&\!\!\equiv\!\!&U_b B\,.
\eea
In general $[A',B'] \ne 0$.  As every experimentalist 
knows, feedback effects often are generated 
inadvertently, particularly in newly-designed experiments, and 
the result is typically\,---\,but not invariably\,---\,noisy 
or chaotic behavior of the device.  The author has conducted 
numerical experiments which support this expectation.  
Depending on the particular choice of $U_a$ and $U_b$, plots 
of the macroscopic variable $q$ and the microscopic variable 
$z$ show a variety of behaviors, ranging from seemingly 
complete randomness to complex trajectories reminiscent of 
chaotic attractors.  However, $U_a$ and $U_b$ together span 
a six-dimensional parameter space which is too large for 
efficient empirical exploration.  We therefore close by 
noting that Eqs.~\ref{eq:interaction}--\ref{eq:prob} provide 
a well-posed framework for investigating noisy and chaotic 
behaviour in the context of quantum measurement theory.  
From a practical point of view, the theory of noisy and 
chaotic devices is surely even more interesting than 
the theory of well-behaved devices.

\bibliographystyle{plain}
\bibliography{MRFMbiblio}

\section*{Acknowledgments}
This research was supported by the University of Washington 
Department of Orthop{\ae}dics, the NIH Biomedical Research 
Technology Area, the Army Research Office, and the NSF 
Instrument Development Program.

\end{document}